\documentclass[journal=jacsat,manuscript=article]{achemso}

\usepackage{chemformula} 
\usepackage[T1]{fontenc} 
\usepackage{graphicx} 
\usepackage{xcolor}

\newcommand{\kp}{\textit{k.p}}
\newcommand{\GeSn}[2]{Ge$_{#1}$Sn$_{#2}$}
\author{Mahmoud R. M. Atalla, Simone Assali, G\'erard Daligou, Anis Attiaoui, Sebastian Koelling, Patrick Daoust, and Oussama Moutanabbir}
\email{ oussama.moutanabbir@polymtl.ca} 
\affiliation{Department of Engineering Physics, \'Ecole Polytechnique de Montr\'eal, C.P. 6079, Succ. Centre-Ville, Montr\'eal, Qu\'ebec, Canada H3C 3A7}
\title {Continuous-wave GeSn light emitting diodes on silicon with $2.5 \, \mu$m room-temperature emission}

\begin{document}







\begin{abstract}
Silicon-compatible short- and mid-wave infrared emitters are highly sought-after for on-chip monolithic integration of electronic and photonic circuits to serve a myriad of applications in sensing and communication. To address this longstanding challenge, GeSn semiconductors have been proposed as versatile building blocks for silicon-integrated optoelectronic devices. In this regard, this work demonstrates light-emitting diodes (LEDs) consisting of a vertical PIN double heterostructure  p-Ge$_{0.94}$Sn$_{0.06}$/i-Ge$_{0.91}$Sn$_{0.09}$/n-Ge$_{0.95}$Sn$_{0.05}$ grown epitaxially on a silicon wafer using germanium interlayer and multiple GeSn buffer layers. The emission from these GeSn LEDs at variable diameters in the 40-120 $\mu$m range is investigated under both DC and AC operation modes. The fabricated LEDs exhibit a room temperature emission in the extended short-wave range centered around 2.5 $\mu$m under an injected current density as low as 45 A/cm$^2$.  By comparing the photoluminescence and electroluminescence signals, it is demonstrated that the LED emission wavelength is not affected by the device fabrication process or heating during the LED operation. Moreover, the measured optical power was found to increase monotonically as the duty cycle increases indicating that the DC operation yields the highest achievable optical power. The LED emission profile and bandwidth are also presented and discussed. 
\end{abstract}


\section{Introduction}

Monolithic infrared (IR) solid-state light sources grown on silicon have attracted significant interest owing to their relevance to scalable photonic integrated circuits (PICs) and compatibility with complementary metal-oxide-semiconductor (CMOS) technologies \cite{powell2022integrated,kim2023short,hu2017silicon}. Among a plethora of applications, this monolithic integration is increasingly crucial to implementing compact and cost-effective IR sensing and imaging technologies. The former requires emitters operating at longer wavelengths in the IR range to overlap with the molecular fingerprint. For instance, emitters in the  $2.3$ $-$ $2.7$ $\mu$m range are useful for sensing both temperature and gas molecules such as  CO, CO$_{2}$, NH$_{3}$, and N$_{2}$O \cite{zia2019high,mathews2021high,stritzke2015tdlas,ji2022mid12,lamoureux2021situ,bader2020progress,ryczko2021interband,olafsen2020optically,dolores2017waveguide,aziz2017multispectral,shterengas2016cascade,li2018dual}.   
These applications call, however,  for broadband emitters such as light emitting diodes (LEDs),  particularly for spectroscopic detection of these critical gases generated in a chemical plant or for their atmospheric monitoring \cite{mikhailova2007optoelectronic,stoyanov2012middle}.
The current emitters serving this wavelength range consist predominantly of III-V compound semiconductors including GaSb- and InP-based device structures \cite{dai2022growth,alexandrov2002portable,sprengel2015continuous,meyer2020interband}. For instance, GaSb LEDs provide wavelength emission peaks between $2.37$ $\mu$m and $2.70$ $\mu$m \cite{danilova2005light,tournie2019mid}. Although GaSb-based lasers and super-luminescent LEDs possess higher tunability of the emitted wavelength \cite{coldren2000monolithic,lee2007widely,sprengel2015continuous,karioja2017multi}, their narrow band emission remains less favorable for spectroscopic gas detection. Notwithstanding the progress in developing III-V LEDs, these materials remain costly and their direct growth on silicon is typically associated with a degradation of device performance \cite{Moutanabbir2010}, thus hindering their integration in large-scale applications. 

 As an alternative, group IV GeSn semiconductors have been explored to circumvent these challenges. These semiconductors are epitaxially grown on silicon and thus can benefit from established scalable manufacturing. Moreover, the Sn content and strain can be controlled to tune the bandgap over the entire range of the IR wavelengths \cite{atalla2023extended,Moutanabbir2021,VondenDriesch2015,Buca2022,wu2023ge}. Up-to-date, the very few reported GeSn LEDs are shown to function in the alternating current (AC) operation mode\cite{Huang2019,Peng2020,Stange2017,bertrand2019mid,Gallagher2015}. While AC-driven LEDs provide quasi-continuous-wave or pulse emission and mitigate the device heating by reducing the thermal power dissipation, direct current- (DC-) driven IR LEDs provide a continuous-wave emission, which is ideal for imaging and portable infrared systems that utilize DC batteries  \cite{thirumalai2018light,lasance2014thermal}. Herein, a GeSn vertical PIN LED with an emission peak in the range of $2.45$ $-$ $2.58$ $\mu$m is demonstrated. The achieved GeSn LEDs are shown to operate under DC bias at a relatively low injection current. Before delving into the device performance, the material growth and characterization are first presented to elucidate the basic properties of the PIN heterostructure including lattice strain and Sn content. Additionally, an eight-band $k.p$ model is implemented to estimate the band alignment of the grown GeSn stack to evaluate the electronic structure of the obtained double heterostructure. Afterward, the electrical and optical measurements are presented to investigate the LED operation under DC- and AC-driven operation modes, the emission wavelength, the emission power profile, and the bandwidth.

\bigskip

\section{Results and discussion}

\label{sec:Results}
\begin{figure}[htbp]
\centering
\includegraphics[width=16cm, height=10.67cm]{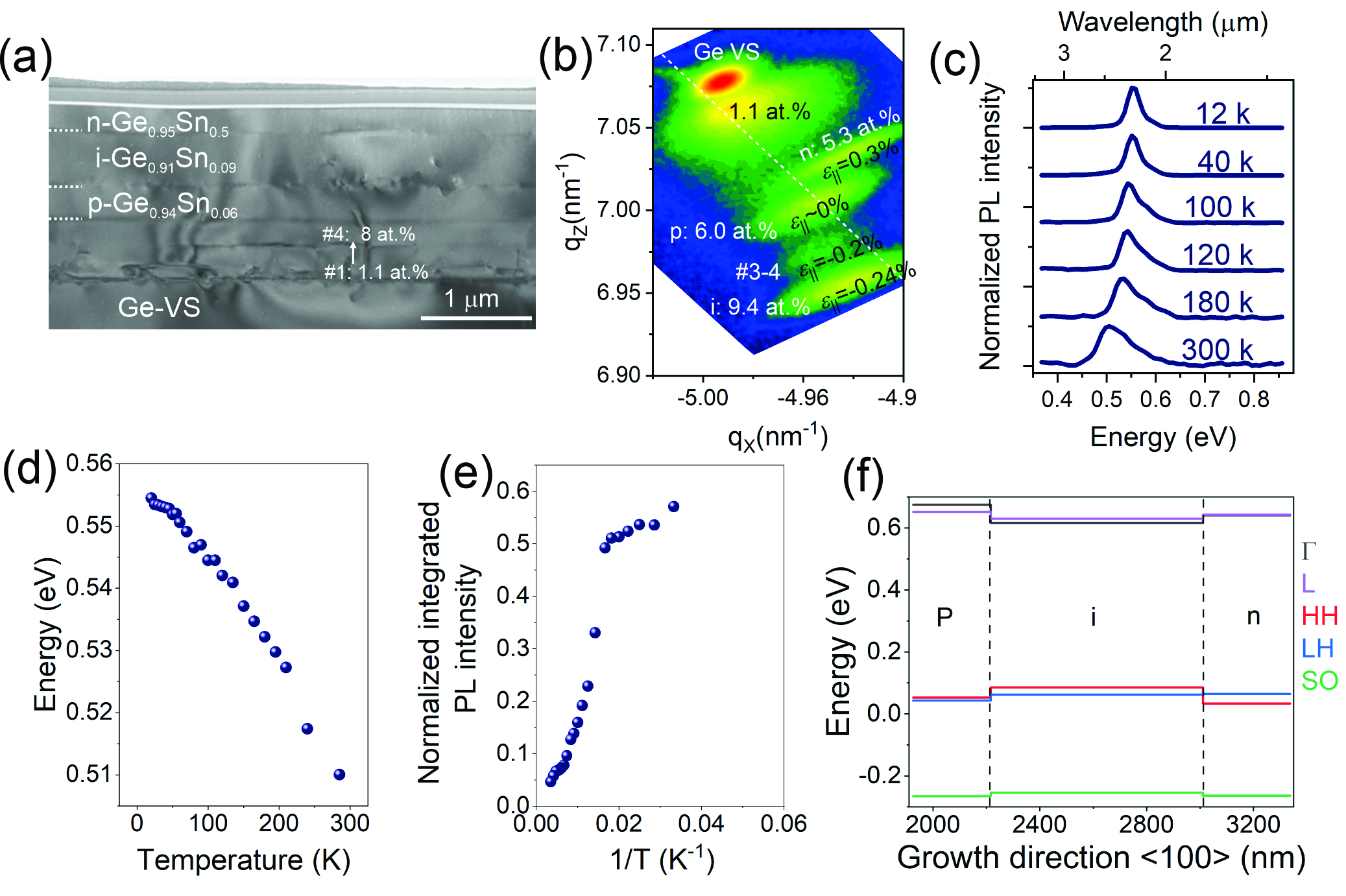}
\caption{GeSn PIN LED basic material properties. (a) TEM micrograph showing the GeSn stack grown on Ge-VS. (b) XRD RSM map around the asymmetrical (-2-24) reflection confirming the Sn composition and strain in different layers. (c) Temperature-dependent PL spectra. (d) The PL energy peak as a function of temperature. (e) Normalized integrated PL intensity as a function of the inverse of temperature. (f) Band lineup at 300K for the GeSn PIN heterostructure calculated using the 8-band $k.p$ method}
\label{fig:Fig__1}
\end{figure}

\indent Starting from a 4-inch Si(100) wafer, a PIN GeSn double heterostructure was grown on a
Ge virtual substrate (Ge-VS) using four GeSn buffer layers in a low-pressure chemical vapor deposition (CVD) reactor. The PIN GeSn layers were grown at temperatures of $335$, $305$, and $345$ $^\circ$C, respectively. The growth temperature was optimized to reach the targeted content and strain while considering the possible fluctuations in Sn incorporation that occur due to the partial relaxation of strain and/or the presence of dopants \cite{assali2019,atalla2022}. The p- and n-type doping were achieved using diborane (B$_{2}$H$_{6}$) and arsine (AsH$_{3}$) precursors. The obtained active doping surpasses $1$ $\times \, 10^{19}$ cm$^{-3}$ for both doping types, as confirmed by capacitance-voltage (C-V) measurements. A cross-sectional transmission electron micrograph (TEM) of the as-grown heterostructure is depicted in Fig.\ref{fig:Fig__1}(a) indicating a total thickness of 1.9 $\mu$m from the Si/Ge-VS interface up to the sample surface. To estimate the strain and Sn composition in this stack, X-ray diffraction (XRD) reciprocal space mapping (RSM) around the asymmetrical (-2-24) reflection was acquired, as displayed in Fig.\ref{fig:Fig__1}(b). The Ge-VS peak indicates an in-plane tensile strain of $\sim 0.2$\%  due to cyclic thermal annealing applied to improve the Ge-VS crystalline quality. However, GeSn buffer layers \#1 to \#4 are under a compressive strain, which remains relatively low reaching $\sim -0.2$\% in \#3 and \#4 layers. The p-Ge$_{0.94}$Sn$_{0.06}$ layer is fully relaxed, whereas the i-Ge$_{0.91}$Sn$_{0.09}$ layer shows a compressive strain of  $-0.24 \%$. Additionally, the pseudomorphic growth on the i-Ge$_{0.91}$Sn$_{0.09}$ yields a tensile strain of $0.3 \%$ in the n-Ge$_{0.95}$Sn$_{0.05}$.  As it can be seen in Fig.\ref{fig:Fig__1}(a), the thicknesses of the n-Ge$_{0.95}$Sn$_{0.05}$, i-Ge$_{0.91}$Sn$_{0.09}$, and p-Ge$_{0.94}$Sn$_{0.06}$ layers are $242$, $432$, and $346$ nm, respectively. While i- and n-layers exhibit high crystalline quality, at the TEM scale, the p-layer and the multi-layer stack underneath show dislocations and defects, especially, at the Si/Ge-VS interface, where the gliding of misfit dislocations is promoted rather than the propagation of threading dislocations toward the upper portion of the stacking.

\bigskip

\bigskip

To investigate the IR emission of the as-grown PIN multi-layer structure, temperature-dependent photoluminescence (PL) measurements were carried out, as shown in  Fig. 1(c). The PL spectra were measured at temperatures in the 12--300 K range. The emission peak is observed at 0.554 eV with a full width at half maximum (FWHM) of 31.8 meV at 12 K. Since the PL emission at 12 K is obtained at a low excitation power density of 231 W/cm$^2$, the constant peak energy from 12 K till 55 K (Fig. 1(d)) is most likely attributed to shallow optically active localized levels near the bandgap edges \cite{assali2021midinfrared}.
However, as the temperature increases above 55 K a progressive redshift is observed up to 300 K, as shown in Fig. 1(d).  In Fig.1(e), the integrated PL intensity, normalized to that measured at 12 K, is plotted as a function of the inverse of temperature. It is noticeable that the PL emission decreases as the temperature increases, and this drop in PL is more remarkable above 55 K. It can be inferred from the monotonic increase of both PL integrated intensity (Fig.1(e)) and emission energy (Fig.1(d)) that, as the temperature decreases, the recombination is dominated by radiative mechanisms rather than phonon and impurity-ionization related nonradiative mechanisms, which is consistent with the band gap directness of the investigated heterostructure over the temperature range 12 K to 300 K \cite{assali2021midinfrared}.


\bigskip
Since both strain and composition can have a significant influence on the GeSn bandgap energy, an eight-band $k\cdot p$ model with envelope function approximation \cite{Bahder1992} was implemented to calculate the band structure of the whole stack. The bandgap lineup in Fig.1(f) is calculated at 300 K. The eight-band \kp{} \GeSn{1-x}{x} material parametrization is based on that reported in early studies \cite{Chang2010,KL2012,Polak_2017}, while strain implementation is based on the Bir-Pikus formalism \cite{PikusBir}. To account for the inaccuracy of Vegard's law in estimating the bandgaps of \GeSn{1-x}{x} alloys, bandgap bowing parameters are introduced for L and $\Gamma$ high-symmetry points \cite{Bertrand2019, Chang2010}. The computed band alignment is displayed in Fig.\ref{fig:Fig__1}(f). It is noteworthy that the i-Ge$_{0.91}$Sn$_{0.09}$ layer has a smaller direct band gap of $0.528$ eV compared to the p-Ge$_{0.94}$Sn$_{0.06}$ and n-Ge$_{0.95}$Sn$_{0.05}$ layers which have bandgap energies of $0.582$ eV and $0.568$ eV, respectively. This result confirms the formation of a double heterostructure at the i-layer of the GeSn PIN stack. It is also noted that tensile strain in the n-layer has reduced its bandgap energy compared to that of the p-layer which contains more Sn content but a higher compressive strain.

\bigskip

\begin{figure}[htbp]
\centering
\includegraphics[width=14.25cm, height=10.5cm]{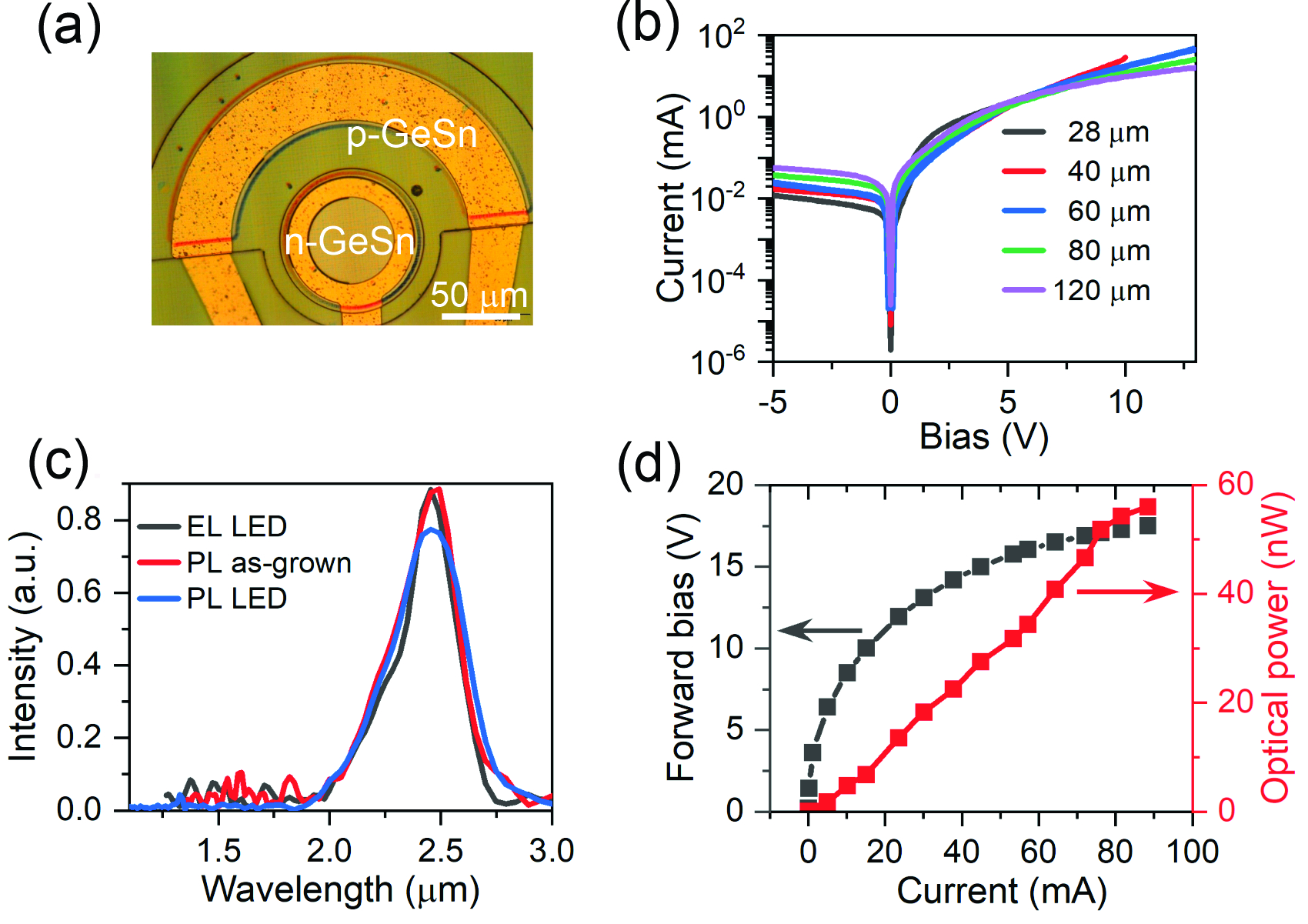}
\caption{GeSn PIN LED fabrication and operation under DC bias. (a) Micrograph of the fabricated GeSn LED device. (b) I-V curves of LEDs with various diameters in the $28-120 \, \mu$m. (c) PL of as-grown and fabricated device compared to the EL from the same $80 \, \mu$m LED device. (d) I-V curves for the DC operation of a representative $80 \, \mu$m GeSn LED}
\label{fig:Fig__2}
\end{figure}

\bigskip

The main steps of the GeSn LED device fabrication include chlorine-based ICP dry etch down to the Ge-VS layer, etch down to the p-GeSn layer forming active circular bumps of various diameters, wet chemical passivation of the etched sidewalls followed by SiO$_2$ deposition, BOE etch of contacts openings in the SiO$_2$ thin-film, and, finally, e-beam deposition of Ti/Au contacts, as shown in Fig.2(a). 
To analyze the LED device performance under DC operation mode, the I-V curves were first measured under DC bias, as shown in Fig.2(b), for device diameters in the range of $28\,\mu$m to $120\,\mu$m. In reverse bias, the measured current ranges from 0.01 mA to 0.08 mA at -5 V as the device's active area increases from $28\,\mu$m to $120\,\mu$m. In forward bias, the LED current reaches about 2 mA at 5 V regardless of the device diameter. This indicates that by simply down-scaling the device size, the rectification ratio increases from approximately one order of magnitude at a device diameter of $120\,\mu$m to two orders of magnitude at a diameter of $28\,\mu$m. As the DC forward bias increases to 10 V, the $40\,\mu$m device reaches breakdown at a current of about 20 mA, while it exceeds 100 mA at a bias $>$ 15 V in larger area devices.

\bigskip

\begin{figure}[htbp]
\centering
\includegraphics[width=16.6cm, height=9.1cm]{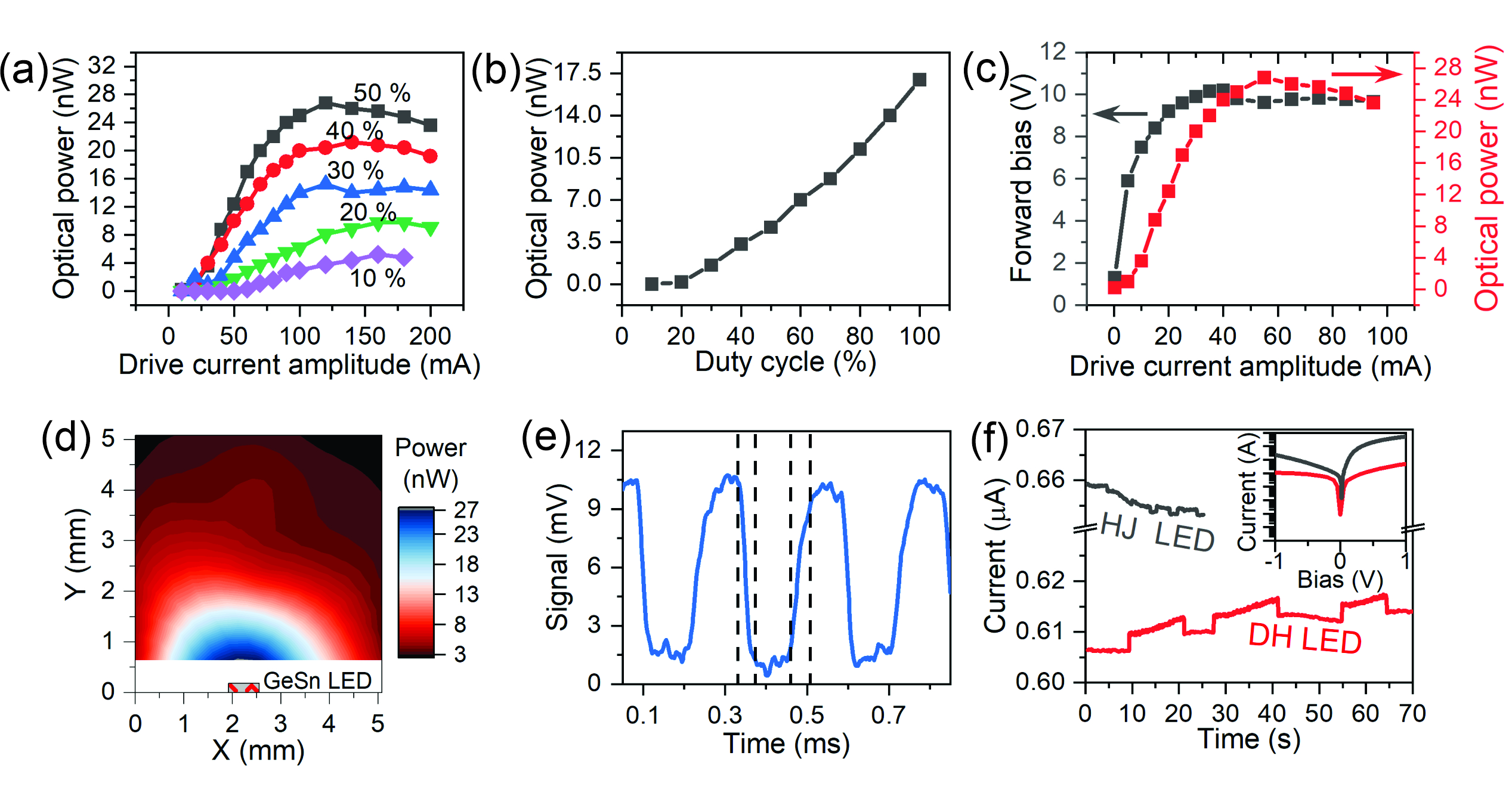}
\caption{LED operation under a quasi-continuous wave (qCW) bias, LED emitted light power profile, and LED bandwidth for $80 \, \mu$m device. (a) Optical power as a function of drive current amplitude in qCW drive mode at various duty cycles. (b) Optical power at a fixed current amplitude of 30 mA but varied duty cycle of the same LED device. (c) Same as (a) but at $50 \, \%$ duty cycle depicting the forward bias as a function of drive current amplitude. (d) Optical power profile under DC drive current of 45 mA. (e) Bandwidth measurement of the GeSn LED. (f) The current of an InGaAs PD measuring the emission of double heterostructure (DH) LED and homojunction (HJ) LED both of the same diameter of $80 \, \mu$m. The inset compares the IV curves of the two LEDs}
\label{fig:Fig__3}
\end{figure}

The electroluminescence (EL) emission of the $80\,\mu$m LED at an injection current of 15 mA was spectrally analyzed using an FTIR spectrometer and liquid nitrogen (LN) cooled InSb photodetector, as displayed in Fig. 2(c). The recorded room temperature EL spectrum peaks at $2.46\,\mu$m. To rule out any shift in the EL peak that might result from any possible stress build-up during device fabrication or heating through the injected current, the EL spectrum was compared to the PL of both an as-grown sample and a fabricated device (Fig.2(c)). It is evident that the EL and PL peaks overlap with each other meaning that the EL peak is intrinsic to the material properties and not affected by the device fabrication process or heating during the LED operation. 

The optical power emitted from the $80\,\mu$m LED was evaluated up to 90 mA yielding 58 nW, as measured by an extended InGaAs (Ex-InGaAs) commercial detector (Fig.2(d)). The emitted optical power is negligible until the DC bias exceeds 3.5 V, which constitutes the turn-on voltage of the LED. While the emitted optical power is nonlinear with the applied DC bias, it is linear with the injection current, which is most likely due to the higher bias required to increase the current injection into the double heterostructure LED. This $80\,\mu$m LED reaches breakdown at 17.8 V. 

The emitted optical power under quasi-continuous wave (qCW) operation mode is investigated, as shown in Fig.3(a).  The qCW is a rectangular voltage waveform applied to the LED with a preset duty cycle and injection current amplitude. As the duty cycle increases from 10\% to 50\%, the emitted optical power increases significantly, however, all curves reach saturation as the amplitude of the injection current exceeds 100 mA most likely because of trap-assisted leakage and non-radiative recombination. Fig.3(b)  shows the emitted optical power as the duty cycle is varied from 10\% to 100\% at an injection current amplitude of 30 mA. The optical power is monotonically increasing as the duty cycle increases indicating that the DC operation of these LEDs would yield the highest achievable optical power. 

The optical power emitted by the $80\,\mu$m LED under qCW bias of 50 \% duty cycle was measured up to an amplitude of 90 mA yielding 28 nW, as shown in Fig.3(c). The emitted optical power is nonlinear with the applied bias and the injection current. This $80\,\mu$m LED reaches saturation around 10.2 V. This saturation is related to the increased injection current above 50 mA equivalent DC current and not related to the applied bias, which remains below the breakdown voltage. 


\bigskip

In Fig.3(d), the optical power profile of the $80\,\mu$m LED is investigated. To acquire this map, the LED was mounted on a scanning stage, and an Ex-InGaAs photodetector was used to measure the emitted optical power as the x position of the LED is varied while the photodetector is moved vertically upward in the y direction after each completed x scan. It is evident that the EL emission follows a Lambertian profile and the emitted power reduces to half as the detector moves 1 mm upward. The highest power of 27 nW was recorded when the detector was closest to the LED (0.7 mm away) and the drive injection current was 45 mA through the $80\,\mu$m LED under DC operation. 

The LED bandwidth is also investigated using the Ex-InGaAs photodetector and a high-bandwidth oscilloscope. The measured signal is displayed in Fig.3(e), and it indicates a fall time of $63.38$ $\mu$s which corresponds to an LED bandwidth of $0.35/$Fall time, i.e. $5.4$ kHz. The low LED bandwidth is most likely related to an increase in the carrier lifetime plausibly caused by the trapped photocarriers at the bandgap energy step at the i-layer and p-layer interface. This was confirmed by comparing the emission of this LED to that of another PIN GeSn LED at a similar Sn content but without a noticeable energy offset at the p- and i-layer interface. The latter exhibits a higher forward bias current compared to the former but emits significantly lower power, as shown in Fig.3(f). Additionally, owing to the relatively small bandgap of the GeSn active layer, thermal radiative emission is expected to be significant,\cite{buca2022room} especially at high driving current, which also reduces the LED bandwidth.

\bigskip
The effect of device diameter size on the emitted power and the change of emission peak position as a function of the driving current is displayed in Fig.4. The EL emission spectra from $40, 80$ and $120\,\mu$m LEDs are measured at various DC injection levels. It is evident that as the device diameter increases, the injection current reduces. Moreover, the lowest injection current, corresponding to the smallest measurable EL, reduces from $159$ A/cm$^2$ ($40\,\mu$m device) to $45$ A/cm$^2$ ($120\,\mu$m device). Another observation from Fig.4 is that the EL peak redshifts as the injection current increases regardless of the device diameter.  This redshift is larger in the $40\,\mu$m device as compared to the $120\,\mu$m device. More precisely, for the $120\,\mu$m device, the EL peak reaches $2.53\,\mu$m at the highest injection current of 500 A/cm$^2$, whereas  the injection current of 5310 A/cm$^2$ in the $40\,\mu$m device corresponds to the EL peak position of $2.58\,\mu$m. The EL peaks are centered at $2.57\,\mu$m and $2.53 \mu$m at DC injection currents of $63.9$ mA and $88.3$ mA in the $40\,\mu$m and $120\,\mu$m LEDs, respectively. The redshift of the EL peak at high current injection levels is most likely caused by the heating effect which becomes more prevalent in smaller devices. 

\bigskip

\begin{figure*}[htbp]
\centering
\includegraphics[width=12cm, height=9cm]{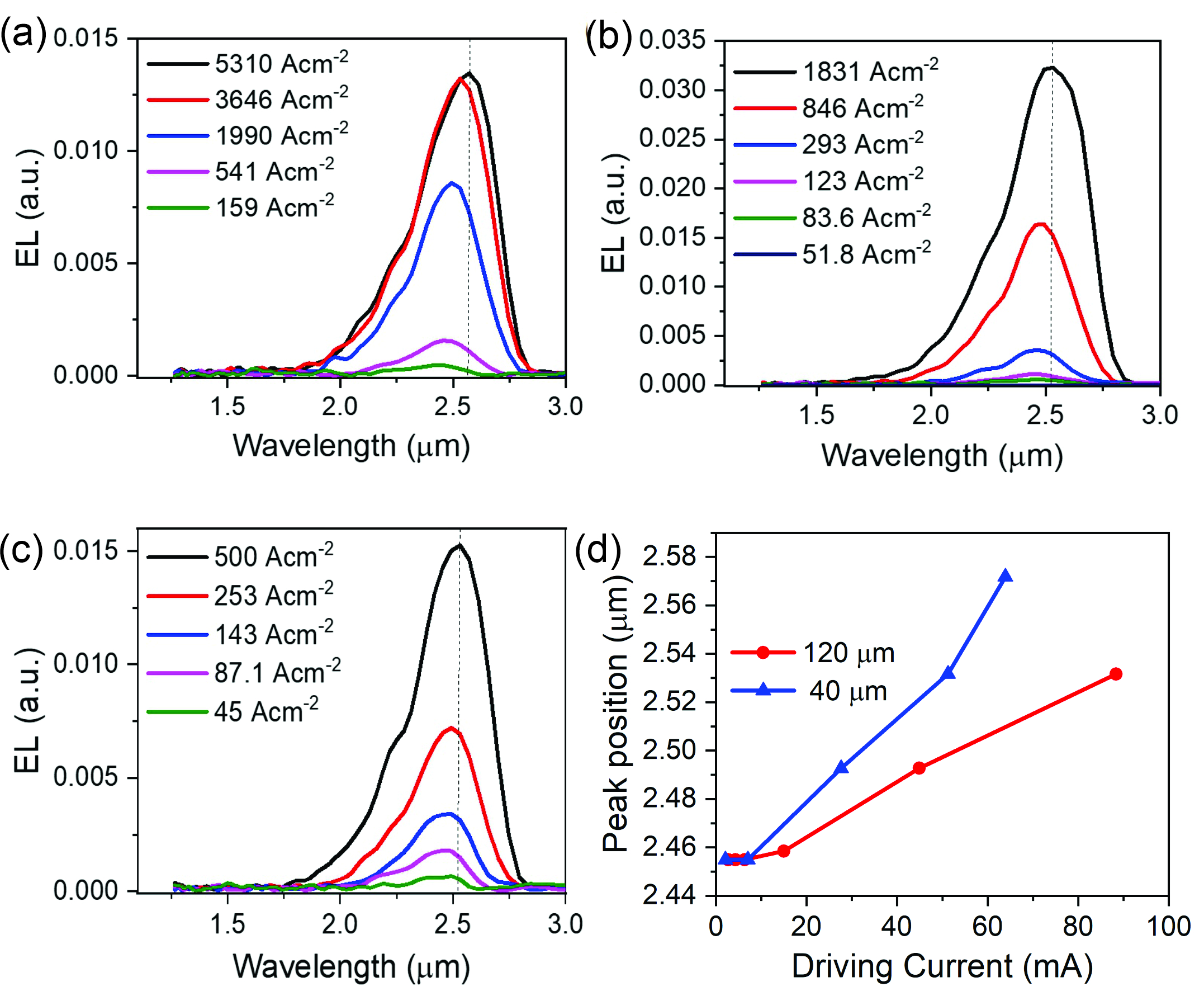}
\caption{Spectral EL as a function of the injection current density
for LED device diameters of $40\,\mu$m (a), $80\,\mu$m (b), and $120\,\mu$m (c). (d) The PL peak position as a function of the DC driving current for two devices of different diameters.}
\label{fig:Fig__4}
\end{figure*}

\section{Conclusion}
In this work, GeSn-based extended-SWIR vertical LEDs monolithically integrated on silicon have been demonstrated. The PIN heterostructure consisting of p-Ge$_{0.94}$Sn$_{0.06}$/i-Ge$_{0.91}$Sn$_{0.09}$/n-Ge$_{0.95}$Sn$_{0.05}$ layers was grown following a step-graded growth protocol to control their strain and content.  The obtained LEDs show an emission peak at a wavelength range of $2.45 - 2.58\,\mu$m depending on the injection current level and the LED device diameter. These LEDs have shown injection current density as low as $45$ Acm$^-2$ for a $120\,\mu$m diameter device and demonstrated both DC and qCW (AC) operation. The emitted power profile was found to be lambartian in shape and the highest emitted power was found to be $58$ nW at $\sim2.57\,\mu$m at an injection current of $100$ mA. These LEDs show a bandwidth that is most likely attributed to an increase in the carrier lifetime due to trapped photocarriers at the i-layer and p-layer interface. The demonstrated group IV GeSn-based LEDs are promising for on-chip integrated extended-IR light emitters for compact and cost-effective gas sensing and imaging applications. However, more work is needed to improve their structural properties and optimize their performance to reach emission powers sufficiently high to be relevant for practical applications.

\bigskip

\subsection{Funding:}
\noindent The work carried out in Montréal was supported by Natural Science and Engineering Research Council of Canada (Discovery, SPG, and CRD Grants), Canada Research Chairs, Canada Foundation for Innovation, Mitacs, PRIMA Québec, Defence Canada (Innovation for Defence Excellence and Security, IDEaS), the European Union's Horizon Europe research and innovation programme under grant agreement No 101070700 (MIRAQLS), and the US Army Research Office Grant No. W911NF-22-1-0277. 
\subsection{Acknowledgement:} The authors would like to thank Joel Bouchard for technical support.

\bibliography{sample.bib}





\end{document}